\documentclass[twocolumn,aps,prl,showpacs,preprintnumbers,superscriptaddress,longbibliography]{revtex4-2}

\usepackage{amssymb}
\usepackage{amsmath}
\usepackage{amsfonts}
\usepackage{dcolumn}     
\usepackage{bm}          
\usepackage{xcolor}
\usepackage{diagbox}
\usepackage{graphicx}    
\usepackage{xcolor}

\usepackage{hyperref}
\hypersetup{
    colorlinks=true,
    linkcolor=blue,
    citecolor=blue,
    urlcolor=blue
  }

\begin{document}
\title{Many-Body Non-Hermitian Physics in the Generalized Brillouin Zone}
\author{Chaoze Lu}
\author{Chuanshu Xu}
\author{Zhenghao Yang}
\author{Xiancong Lu}
 \email{xlu@xmu.edu.cn}
\affiliation{Department of Physics, Xiamen University, Xiamen 361005, China}

\begin{abstract}
  The breakdown of conventional bulk-boundary correspondence (BBC) in
  non-Hermitian system can be resolved by the generalized Brillouin
  zone (GBZ) theory. However, extending the GBZ theory to interacting
  many-body systems remains an open problem. Here, we consider an
  interacting non-Hermitian model characterized by a circular GBZ. We
  show that, based on a GBZ transformation, a quasi-reciprocal
  many-body Hamiltonian can be constructed which, under periodic
  boundary conditions (PBC), captures the physics of the original
  non-Hermitian model under open boundary conditions (OBC).  Using
  exact diagonalization (ED), we determine the phase diagram for the
  quasi-reciprocal many-body Hamiltonian by computing the Zak phase
  and the structure factor of the charge-density-wave (CDW) phase. We
  further investigate the entanglement properties and find that the
  degeneracy of the low-lying entanglement spectrum characterizes each
  phase in the phase diagram. These findings demonstrate that the
  topological properties in interacting non-Hermitian system is encoded in
  the entanglement spectrum of the quasi-reciprocal model. Our work
  establishes a route to studying many-body non-Hermitian physics
  within the GBZ formalism.
\end{abstract}
\maketitle






{\color{blue} \itshape Introduction.} Non-Hermitian Hamiltonians, which describe open systems interacting
with their environments, have attracted significant attention across
various fields of physics
\cite{as.go.20,be.bu.21,zh.zh.22,ok.sa.23,ya.li.24,go.ba.25}.  A
unique phenomenon in non-Hermitian systems is the anomalous
accumulation of bulk eigenstates at the system's boundaries, which is
termed non-Hermitian skin effect
(NHSE)\cite{ya.wa.18,ya.so.18,ku.ed.18,go.as.18,le.th.19,ok.ka.20}.
In the presence of NHSE, the energy spectra under open boundary
conditions (OBC) and periodic boundary conditions (PBC) differ
significantly \cite{lee.16,xion.18}, leading to the breakdown of the
well-known bulk-boundary correspondence (BBC)
\cite{ha.ka.10,qi.zh.11}.  Interestingly, the topology of
non-Hermitian systems can be described by non-Bloch band theory: zero
modes under OBC correspond to bulk topological invariants defined on
the generalized Brillouin zone (GBZ)
\cite{ya.wa.18,yo.mu.19,ya.zh.20}, rather than on the conventional
Brillouin zone (BZ). This breakthrough has sparked immense research
interest in the theory of GBZ
\cite{ya.so.18,zh.ya.22,ji.le.23,wa.so.24,hu.25}.

The skin effect in the context of many-body physics exhibits behavior
distinct from that at the single-particle level.  A variety of novel
phenomena have been investigated and identified, including real space
Fermi surface \cite{mu.le.20}, topological Mott
phases\cite{zh.ch.20,xu.ch.20,li.he.20}, spectral clusters
\cite{zh.de.22,ka.sh.22,lu.su.24,qi.li.24}, Fock space skin
effects \cite{sh.sa.24,sh.qi.24,wa.li.25}, and skin effects of a kink \cite{zh.pa.25}.  Several approaches have
also been proposed to characterize the many-body skin effect
\cite{al.he.22,gl.de.24,ha.ka.25}.  So far, most studies on the
non-Hermitian many-body properties have started from the original
Hamiltonian in real space. Given the great success of GBZ theory in
describing the non-interacting skin effect, a natural question arises:
is it possible to explore the non-Hermitian many-body physics from the
perspective of GBZ theory? This topic, however, remains largely
unexplored \cite{ya.lu.24,wa.wa.24}.

An important aspect of many-body phases and phase transitions is their
entanglement behavior, which can be quantified by entanglement entropy
(EE) \cite{am.fa.08,ei.cr.10}.  Recently, the concept of entanglement
entropy has been generalized to non-Hermitian systems within the
biorthogonal framework \cite{ch.yo.20,he.re.19}.  This has inspired
substantial investigations on non-Hermitian EE
\cite{le.le.20,ch.ch.21,ok.sa.21a,ba.do.21,mo.ma.21,gu.yu.21,ch.pe.22,lee.22,tu.tz.22,ch.zh.22,fo.ar.23,ka.nu.23,xu.li.23,yi.ha.23,pa.wa.23,hs.ch.23,or.im.23,ga.tu.23,li.yu.24,zhou.24,li.li.24,wa.fa.24,ch.zh.24},
although its physical interpretation remains somewhat ambiguous
\cite{tu.tz.22,fo.ar.23,ya.lu.24}.  However, the non-Hermitian
entanglement spectrum, which contains more comprehensive information
than entanglement entropy \cite{li.ha.08}, has received far less
attention \cite{he.re.19,ch.yo.20,sa.yu.22,or.ca.22,pa.ry.25}. In
particular, its relationship to many-body topological phases remains
largely unclear, especially for interacting systems.

In this paper, we fill these gaps by generalizing the GBZ
transformation to interacting non-Hermitian systems. We show that for
the circular GBZ, a quasi-reciprocal lattice Hamiltonian can be
constructed from the original non-Hermitian interacting
model. Remarkably, the properties of the original non-Hermitian model
under OBC can be captured by the quasi-reciprocal Hamiltonian under
PBC. We analyze the entanglement spectrum (ES) of the
quasi-reciprocal Hamiltonian and demonstrate that the degeneracy of ES
serves as a robust indicator for many-body topological phases in
non-Hermitian systems.

\begin{figure}[tb]
\centering
\includegraphics[width=1.0\linewidth]{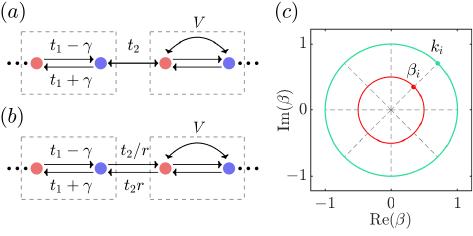}
\caption{Schematic diagram of the non-Hermitian interacting SSH model
  (a) and the corresponding quasi-reciprocal model (b). (c) The
  correspondence between $\beta_i$ in the circular generalized
  Brillouin zone and crystal momentum $k_i$ in the conventional
  Brillouin zone.}
\label{fig:GBZt}
\end{figure}

{\color{blue} \itshape Model and Method.} We study a spinless
non-Hermitian SSH model \cite{ya.wa.18,lieu.18} with nearest-neighbor (NN)
interactions \cite{zh.pa.25,si.ma.14}, described by the Hamiltonian:
\begin{align}
        \mathcal{H}= & \sum_{i=1}^{N}\left[\left(t_1+\gamma\right)c_{2i-1}^{\dagger} c_{2i}
                     + \left(t_1-\gamma\right)c_{2i}^{\dagger}
                       c_{2i-1}\right] \nonumber\\
                   & + \sum_{i=1}^{N}\left[t_2 c_{2i}^{\dagger} c_{2i+1}+t_2
                     c_{2i+1}^{\dagger} c_{2i}\right]
                     + \sum_{j=1}^{L} Vn_jn_{j+1},
        \label{eq:H_SSHV}
\end{align}
where $c_{i}^{\dagger}$($ c_{i}$) is the fermionic creation
(annihilation) operator at the $i$-th site, $t_1$ and $t_2$ are the
intracell and intercell hopping amplitudes, respectively, $\gamma$
denotes the non-reciprocal contribution to the hopping. The hopping
processes are illustrated in Fig.~\ref{fig:GBZt}(a). Here,
$n_j=c_j^{\dagger} c_j$ is particle number operator, $V$ denotes the
nearest-neighbor interaction strength, $N$ is the number of unit
cells, and the total lattice size is $L=2N$. 
In the remainder of this paper, we set $t_2=1$ as the energy unit and
focus on the half-filled system, \textit{i.e.}, the particle number equals $N$.


In non-Bloch band theory, the conventional crystal momentum $k$ is
generalized to complex values $s$. The trajectory of $\beta=e^{is}$
forms a loop on the complex plane, known as GBZ
\cite{ya.wa.18,yo.mu.19}.
We demonstrate that, by imposing an appropriate boundary condition on
the real-space Hamiltonian, the non-Bloch Hamiltonian $h(\beta)$ can
be deduced by a similarity transformation $\mathcal{U}$; see
Eq.~(S6) of the Supplemental Material for an example.
This transformation
$\mathcal{U}$ can also be applied to the interaction term, allowing
the Hamiltonian to be expressed in the non-Bloch basis.  For the
circular GBZ, the modulus $r=|\beta|$ is constant and the argument of
$\beta$ is uniformly distributed along the circle, as shown in
Fig.~\ref{fig:GBZt}c. Thus, there is a correspondence between each
momentum $k_i$ in the conventional BZ and $\beta_i$ in GBZ:
$\beta_i=re^{ik_i}$.  By performing a Fourier transformation
$\mathcal{F}$ with respect to the crystal momentum $k$ on both the
kinetic and interaction terms, an "artificial" quasi-reciprocal
lattice Hamiltonian can be constructed that possesses the same energy
spectrum and topological properties as the original Hamiltonian
\cite{ya.lu.24}.

The quasi-reciprocal Hamiltonian corresponding to the interacting SSH
model in Eq.~(\ref{eq:H_SSHV}) is given by (see Sec.~IB of the
Supplemental Material for a derivation):
\begin{align}
\label{eq:Ht_SSHV}
  \tilde{\mathcal{H}}= & \sum_{i=1}^{N}\left[\left(t_1+\gamma\right)a_{2i-1}^{\dagger} a_{2i}
                         +\left(t_1-\gamma\right)a_{2i}^{\dagger} a_{2i-1}\right]\nonumber\\
                       & +\sum_{i=1}^{N}\left[t_2 r a_{2i}^{\dagger}
                         a_{2i+1}+\frac{t_2}{r} a_{2i+1}^{\dagger} a_{2i}\right]
                         +\sum_{j=1}^{L} V\tilde{n}_j\tilde{n}_{j+1},
\end{align}
where $a_{i}^{\dagger}$($a_{i}$) is the creation (annihilation)
operator at the quasi-reciprocal lattice,
$\tilde{n}_j=a_j^{\dagger}a_j$ is the particle number operator in the
new basis, and
$r=\sqrt{|\left(t_1-\gamma\right)/\left(t_1+\gamma\right)|}$ is the
radius of the GBZ.
The operators $a_{i}^{\dagger}$($a_{i}$) are related to
$c_{i}^{\dagger}$($c_{i}$) by the similarity transformation $\mathcal{S}=\mathcal{U}\mathcal{F}^{\dagger}$:
\begin{align}
\label{eq:cjaj}
c_{i} = \sum_{m} \mathcal{S}_{im}a_m = r^{i}a_{i}, \quad
c_{i}^{\dagger} = \sum_{m} a_m^{\dagger}\mathcal{S}_{mi}^{-1} = r^{-i}a_{i}^{\dagger}.
\end{align}
Here, the index $i$ labels the unit cell, and the $A$ and $B$
sublattices are not explicitly written out.  After the transformation,
the intercell hopping amplitude is renormalized, whereas both the
intracell hopping amplitude and the interaction strength remain
unchanged.
%
%
%
%
%
%
The quasi-reciprocal Hamiltonian $\tilde{\mathcal{H}}$ serves as a
surrogate for the original Hamiltonian $\mathcal{H}$ and yields
exactly the same OBC energy spectrum, as shown in Fig.~S1 of the Supplemental
Material. Moreover, $\tilde{\mathcal{H}}$ is free of the skin effect,
numerically tractable, and much closer to a Hermitian system.
We diagonalize the quasi-reciprocal model using the non-Hermitian
Lanczos method \cite{ba.de.00,ch.pe.22}. The left and right many-body
ground states correspond to the eigenstates with the smallest real
parts of the energy \cite{ch.yo.20,he.re.19,gu.yu.21}.

\begin{figure}[tb]
	\centering
	\includegraphics[width=1.0\linewidth]{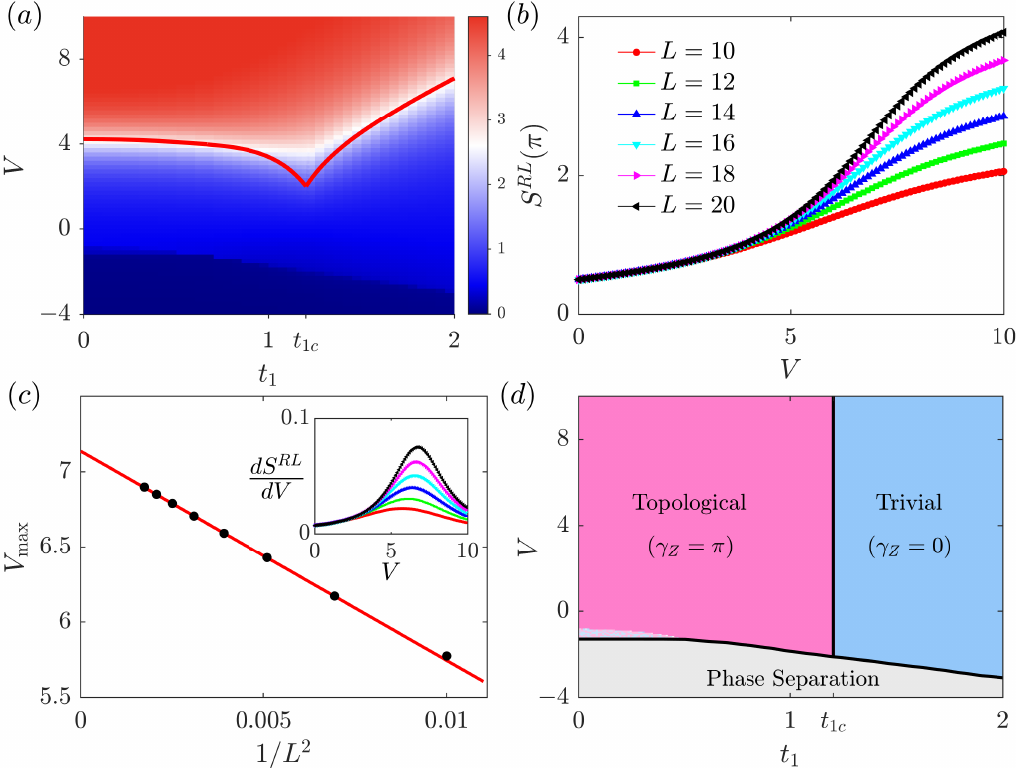}
	\caption{(a) Charge-density-wave structure factor
          $S^{RL}(\pi)$ in the $t_1$–$V$ plane for a fixed lattice
          size $L=20$. (b) $S^{RL}(\pi)$ as a function of the
          interaction $V$ at $t_1=2$ for various lattice sizes
          $L$. (c) Finite-size extrapolation of $V_{\mathrm{max}}$ as
          a function of $1/L^2$, where $V_{\mathrm{max}}$ denotes the
          point at which the derivative $d S^{RL}(\pi)/dV$ is
          maximized. The inset shows the derivative $d S^{RL}(\pi)/dV$
          corresponding to the data in panel (b).  (d) Zak phase
          $\gamma_{Z}$ of $\tilde{\mathcal{H}}$ under PBC. The black
          line indicates the topological phase boundary, with
          $\gamma_{Z} = \pi$ in the topological region and
          $\gamma_{Z} = 0$ in the trivial region.  Other parameters
          are $t_2=1$ and $\gamma=2/3$.}
\label{fig:Scdwt}
\end{figure}

{\color{blue} \itshape Phase diagram.} 
We now investigate the phase diagram of the quasi-reciprocal
Hamiltonian $\tilde{\mathcal{H}}$ under PBC, which reflects the
physics of the original Hamiltonian $\mathcal{H}$ under OBC.
%
%
Similar to the Hermitian SSH model, a Charge Density Wave (CDW) state
emerges as the repulsive interaction $V$ increases
\cite{mi.ca.11,me.ju.23}.
To probe the phase transition to the CDW state, we introduce the
following biorthogonal structure factor,
\begin{align}
\label{equ:SkRL}
   S^{RL}(k)=&\frac{1}{L^2}\sum_{i,j}e^{ik(i-j)}\left( \langle \psi_L|
               \tilde{n}_i\tilde{n}_j | \psi_R \rangle \right.\nonumber\\
            &\left. -\langle \psi_L| \tilde{n}_i | \psi_R\rangle \langle \psi_L | \tilde{n}_j | \psi_R\rangle\right).
\end{align}
where $|\psi_L\rangle$ and $|\psi_R\rangle$ denote the left and right
eigenstates of $\tilde{H}_{\mathrm{PBC}}$, respectively.  The
biorthogonal structure factor $S^{RL}(k)$ is well defined, as it is
always real-valued. Moreover, in large $V$ regime, $S^{RL}(k)$
exhibits a sharp peak at $k=\pi$, consistent with the density profile
of the original Hamiltonian under OBC; see, for example, Fig.~S3 in
the Supplemental Material. In contrast, the structure factor
$S^{RR}(k)$, defined solely in terms of the right states, remains
nearly zero for large $V$; see Fig.~S4 in the Supplemental Material
for illustration. These observations justify the use of
$S^{RL}(k)$ as an appropriate diagnostic of the CDW state.  The value
of $S^{RL}(\pi)$ increases with increasing interaction strength $V$
and becomes large in the strong-coupling regime, signaling the onset
of CDW order, as shown in Fig.~\ref{fig:Scdwt}(a) and (b).  We use the
maximum of the derivative $dS(\pi)/dV$ to identify the phase
transition to the CDW state. Specifically, we extract the values
$V_{\max}$ corresponding to the peaks of $dS(\pi)/dV$ for different
system sizes $L$, and then extrapolate them to the thermodynamic limit
to obtain the critical interaction strength $V_{c}$. One
representative example is shown in Fig.~\ref{fig:Scdwt}(c).

\begin{figure}[htb]
	\centering
	\includegraphics[width=0.84\linewidth]{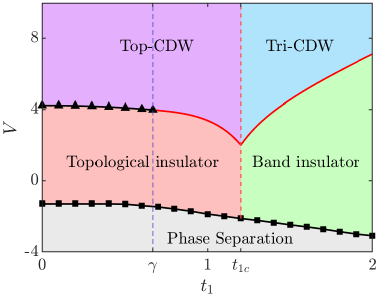}
	\caption{ Phase diagram of the quasi-reciprocal SSH model in the
          $t_1-V$ plane. The Top-CDW and Tri-CDW represent topological
          and trivial CDW phases, respectively.  The red line shows
          the fitted CDW phase boundary given in the main text, while
          the black line with triangles denotes the phase boundary
          obtained from numerical extrapolation.  The black
          line with squares indicates the phase-separation boundary
          obtained numerically for $L=20$.  The red dashed line
          ($t_{1c}\approx 1.2$ ) marks the
          topological phase boundary, and the blue dashed line
          ($t_1=\gamma$) indicates the non-Hermitian exceptional
          point. For $t_1<\gamma$, the energy spectrum becomes
          complex. Other parameter: $t_2=1$ and $\gamma=2/3$.}
	\label{fig:pd}
\end{figure}






The extrapolated critical interaction strength $V_c$ for
$0 < t_1 < \gamma$ is shown in Fig. \ref{fig:pd} (black
triangles). For $t_1 > \gamma$, the CDW phase boundary (red line) is
determined by a fitting procedure, as explained in Sec. IV of the
Supplemental Material.  When $t_{1}>\gamma$, the quasi-reciprocal
model $\tilde{\mathcal{H}}$ can be mapped onto a Hermitian SSH model
$\bar{\mathcal{H}}$ via the similarity transformation
$\mathcal{S}'=\mathrm{diag}\{1/r,1,1/r,1,\dots,1/r,1\}$
\cite{ya.lu.24}.
Since the structure factor $S^{RL}(\pi)$ of
$\tilde{\mathcal{H}}_{\mathrm{PBC}}$ is identical to that of
$\bar{\mathcal{H}}_{\mathrm{PBC}}$, the CDW phase boundaries of
$\tilde{\mathcal{H}}$ and $\bar{\mathcal{H}}$ coincide. For the
Hermitian SSH model, the CDW phase boundary is obtained by fitting ED
data in the regime where the correlation length is much smaller than
the system size $(\xi\ll L)$ to the analytical result at the critical
point; see Sec. IV.A of the Supplemental Material for details.  We
firstly determine the CDW phase boundary $V_c$ of the
Hermitian $\bar{\mathcal{H}}$ and then obtain the phase boundary of
$\tilde{\mathcal{H}}$ by substituting the form of renormalized
intercell hopping. 
The fitted phase boundary is given by
$V_c=4-\frac{1}{50}10^{2\bar{t}_1}$ for $\gamma<t_1<t_{1c}$, and
$V_c=\bar{t}_1(4-\frac{1}{50}10^{2/\bar{t}_1})$ for $t_1>t_{1c}$, where
$\bar{t}_1 = \sqrt{(t_1 - \gamma)(t_1 + \gamma)}$ is the intercell
hopping of $\bar{\mathcal{H}}$.

The topological properties of an interacting system can be determined
from the Zak phase computed under twisted boundary conditions \cite{yu.li.16,me.ju.23}.
For non-Hermitian systems, we adopt the biorthogonal Zak phase defined as
\begin{equation}
	\gamma_{Z} = i \int_0^{2\pi}d\theta
        \left\langle\psi_{L}(\theta)|\partial_{\theta}|
          \psi_{R}(\theta)\right\rangle,
	\label{eq:Zak}
\end{equation}
where $\theta$ denotes the twist angle imposed on the boundary hopping
terms.  The topological phase diagram of the quasi-reciprocal model
$\tilde{\mathcal{H}}$ is shown in Fig. \ref{fig:Scdwt}(d). For
$V = 0$, the topological transition occurs at
$t_{1c} = \sqrt{t_2^2 + \gamma^2}$, consistent with the edge states of
the original $\mathcal{H}$ under OBC. Remarkably, the phase boundary
remains unchanged as $V$ increases.




According to the CDW structure factor and the Zak phase, five phases
can be identified in the phase diagram shown in Fig. \ref{fig:pd}. For
small $V$, the system is either in a topological insulator phase
($t_1 < t_{1c}$) or a trivial band insulator phase ($t_1 >
t_{1c}$). As $V$ increases, both phases develop CDW order. In the
large-$V$ regime, the system evolves into a topological CDW phase for
$t_1 < t_{1c}$ and a trivial CDW phase for $t_1 > t_{1c}$.  Finally,
for sufficiently strong attractive interaction ($V < 0$), the system
undergoes phase separation, with particles clustering together.
In this regime, the structure factor is nearly zero [see
Fig.~\ref{fig:Scdwt}(a)], and the Zak phase becomes ill-defined due to
the high degeneracy.

\begin{figure}[htb]
	\centering
	\includegraphics[width=1.0\linewidth]{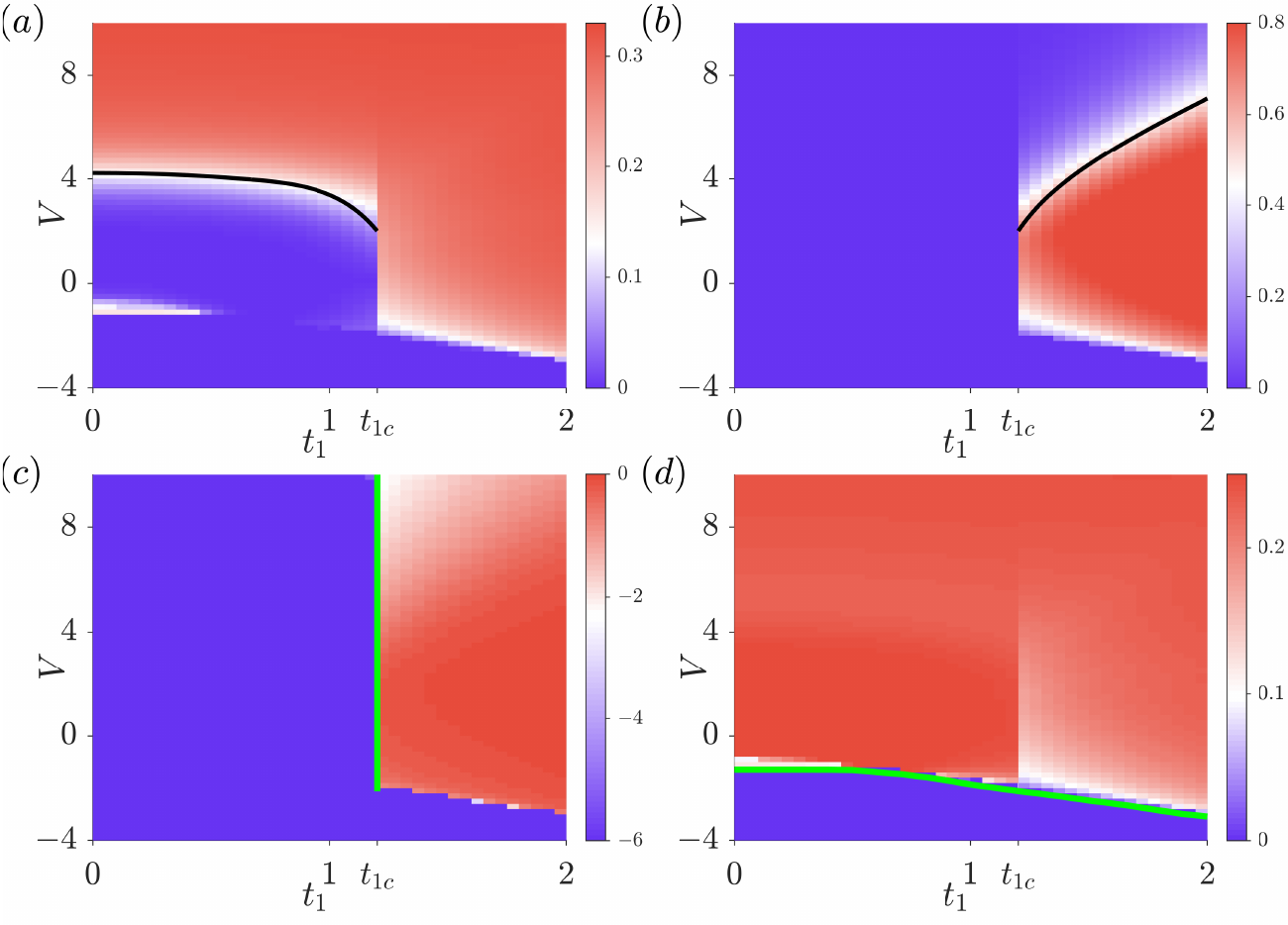}
	\caption{Degeneracy structure $D(n)$ of the entanglement
          spectrum of the quasi-reciprocal Hamiltonian
          $\tilde{\mathcal{H}}$ in the $t_1$–$V$ plane.  (a)
          $D(4)$. (b) $D(2)$. (c) $\log_{10}[D(2)+10^{-6}]$, where
          $10^{-6}$ is introduced to avoid the divergence of
          $\log_{10}(0)$. (d) $D(5)$. The black and green lines
          represent the phase boundary in Fig. \ref{fig:pd}.  Common
          parameters: $L=20$, $N=10$, $t_2=1$, and $\gamma=2/3$.}
	\label{fig:DrhoE}
      \end{figure}

{\color{blue} \itshape Degenercy of Entanglement Spectrum.}
Entanglement is a fundamental feature of many-body quantum states.
As noted in the well-known Li–Haldane conjecture \cite{li.ha.08}, the
low-lying entanglement spectrum can serve as a “fingerprint” to identify
topological order.
Whether this conjecture applies to non-Hermitian many-body systems
remains unclear. Here, we analyze the low-lying degeneracy of the
entanglement spectrum (ES) of the quasi-reciprocal Hamiltonian
$\tilde{\mathcal{H}}$ within the GBZ framework and explore its
connection to the ground-state degeneracy of the original
$\mathcal{H}$ under OBC \cite{yu.ya.24,si.ma.14,ye.mu.16}.

The biorthogonal density matrix is defined as
$\rho^{RL} = |\psi_R\rangle \langle \psi_L|$. Upon partitioning the
system into two subsystems $A$ and $B$, the reduced density matrix of
subsystem $A$ is obtained by tracing out all degrees of freedom in
subsystem $B$ \cite{he.re.19,ch.yo.20,gu.yu.21,ya.lu.24}, \textit{i.e.},
$\rho^{RL}_A = \mathrm{Tr}_B \rho^{RL}$.
The entanglement spectrum $\xi_l$ is the eigenvalues of the $\rho^{RL}_{A}$,
and the entanglement entropy is given by $S_A = - \mathrm{Tr} \left[
\rho^{RL}_A \log (\rho^{RL}_A) \right]$.
%
%
To quantify the low-order degeneracy of the entanglement spectrum
$\{\xi_l\}$ arranged in ascending order, we introduce the $n$th-order
degeneracy $D(n)$ \cite{si.ma.14}, defined as
\begin{equation}
	D(n)=\left[ (n-1)\xi_n-\sum_{i=1}^{n-1}\xi_i\right] / (n-1).
	\label{eq:Dn}
\end{equation}
A vanishing $D(n)$ implies that the system exhibits at least $n$-fold
degeneracy, whereas a sharp change in $D(n)$ might indicate a phase
transition.

The degeneracy of the entanglement spectrum of the quasi-reciprocal
Hamiltonian $\tilde{\mathcal{H}}$ under PBC is presented in
Fig.~\ref{fig:DrhoE}.  In the topological insulator phase
($t_1<t_{1c}$), $D(4)$ is nearly zero [see Fig.~\ref{fig:DrhoE}(a)],
indicating a fourfold degeneracy of the entanglement spectrum. By
contrast, in the band-insulator phase ($t_1>t_{1c}$), $D(2)$ is finite
[see Fig.~\ref{fig:DrhoE}(b)], indicating a nondegenerate entanglement
spectrum. These behaviors are similar to those in the Hermitian SSH
model \cite{si.ma.14,ye.mu.16}.  As the interaction $V$ increases,
both the topological-insulator and the band-insulator phases evolve
into the CDW phase. A sharp change in the degeneracy occurs at the CDW
phase boundary [the black lines in Fig.~\ref{fig:DrhoE}(a) and
(b)]. Notably, two types of CDW phases appear, distinguished by their
degrees of degeneracy.  In the topological-CDW phase, the low-lying
entanglement spectrum is exactly twofold degenerate, reflecting the
underlying topological nature of the phase.  In contrast, in the
trivial-CDW phase the spectrum is only approximately twofold
degenerate: $D(2)$ remains a small but finite for intermediate $V$ and
vanishes only in the limit $V \to \infty$.  The degeneracy in this
phase originates from the approximate twofold degeneracy of the ground
state under OBC, namely, the degeneracy between the states
$|101010\dots\rangle$ and $|010101\dots\rangle$.  Particularly, we
plot $\log_{10}[D(2)+10^{-6}]$ in Fig.~\ref{fig:DrhoE}(c). It is clear
that $D(2)$ is nearly zero (smaller than $10^{-6}$) in topological
phases, whereas it is larger than $10^{-6}$ in the trivial-CDW phase.
The logarithm of the degeneracy $\log_{10}[D(2)]$ exhibits a quantized
structure that clearly delineates the topological phase boundary at
$t_{1c}$, in agreement with the Zak phase shown in
Fig.~\ref{fig:Scdwt}(d).  For sufficiently strong attractive
interaction $V<0$, the entanglement spectrum becomes highly
degenerate. We plot $D(5)$ in Fig.~\ref{fig:DrhoE}(d), which clearly
distinguishes the phase-separated region from the other, less
degenerate phases in the phase diagram.  Overall, the degeneracy of
the entanglement spectrum provides a unified and complete
characterization of the phase diagram in Fig.~\ref{fig:pd}.
Importantly, this method is largely insensitive to the lattice size
and allows one to extract information at a lower computational cost.

{\color{blue} \itshape Conclusion and Discussion.}
We study a non-Hermitian SSH model with nearest-neighbor interactions
within the GBZ formalism. For a circular GBZ, a one-to-one
correspondence exists between the conventional crystal momentum and
the GBZ momentum. Exploiting this relation, we generalize the GBZ
similarity transformation to interacting systems and construct a
quasi-reciprocal many-body Hamiltonian.  Five phases are identified in
the phase diagram of the quasi-reciprocal model. The
charge-density-wave phase boundary is determined by finite-size
extrapolation combined with a fitting procedure, while the topological
phase boundary is identified from the Zak phase. Interestingly, we
demonstrate that the degeneracy of low-lying ES is distinct across the
five phases, establishing it as a robust indicator of topological
phases in non-Hermitian systems.
The quasi-reciprocal model, being
closer to a Hermitian system, provides an excellent surrogate for the
original non-Hermitian model.




When the GBZ is noncircular, its radius is no longer constant and the
GBZ momenta do not coincide with the conventional crystal
momenta. After performing the GBZ transformation, the hopping and
interaction terms of the quasi-reciprocal model inevitably acquire
long-range contributions \cite{ya.lu.24,le.li.20}. The model
parameters depend on the structure of the GBZ, rendering the model
highly complex.




\begin{acknowledgments}
  We are grateful for the discussions with Zhesen Yang and Shijie
  Hu. This work is supported by the National Natural Science
  Foundation of China (Grants No. 12574162 and No. 11974293) and the
  Natural Science Foundation of Xiamen, China (Grants
  No. 3502Z202473008).
\end{acknowledgments}

\bibliographystyle{apsrev4-1}
\bibliography{ref}

\end{document}


\title{Supplemental Material for: \\
  Many-body non-Hermitian Physics in the Generalized Brillouin Zone}
\author{Chaoze Lu}
\author{Chuanshu Xu}
\author{Zhenghao Yang}
\author{Xiancong Lu}
 \email{xlu@xmu.edu.cn}
\affiliation{Department of Physics, Xiamen University, Xiamen 361005, China}







\maketitle


\section{Quasi-reciprocal Hamiltonian}
\label{App:C}

\subsection{Similarity Transformation in GBZ theory}

We first consider the quadratic part of a non-Hermitian Hamiltonian
$\mathcal{H}=\sum_{i,j}c^\dagger_iH_{ij}c_j$, where
$\mathcal{H}\neq \mathcal{H}^\dagger$ and $c_i^\dagger (c_i)$ is the
fermionic creation (annihilation) operator on lattice site $i$.
Under periodic boundary condition, the Hamiltonian $\mathcal{H}$ can
be diagonalized as $\mathcal{H}=\sum_{k}h(k)c_k^{\dagger}c_k$ by a
unitary Fourier transformation $\mathcal{F}$,
\begin{align}
\label{eq:cjck}
c_j &= \sum_{k} \mathcal{F}_{jk}c_k ,\quad  \mathcal{F}_{jk} = \langle
      j | k\rangle = \frac{1}{\sqrt{N}}  e^{i k j} \nonumber\\
c_j^{\dagger} &= \sum_{k} c_k^{\dagger} \mathcal{F}_{kj}^{\dagger} ,\quad  \mathcal{F}_{kj}^{\dagger} = \langle
      k | j\rangle = \frac{1}{\sqrt{N}}  e^{-i k j} 
\end{align}
with $k$ being the conventional crystal momentum.
The Bloch Hamiltonian $h(k)$ gives the energy spectrum of
$\mathcal{H}$ under PBC, which is significantly different from the
energy spectrum under OBC.
This discrepancy can be remedied by the GBZ theory.  The non-Bloch
Hamiltonian $h(\beta)$ is obtained from $h(k)$ by replacing $e^{ik}$
with $\beta=e^{is}$, where $s$ is a complex number \cite{ya.wa.18,yo.mu.19}.  The
$h(\beta)$ along the GBZ trajectory reproduces the OBC energy spectrum
in the thermodynamic limit.

The assumption of GBZ theory implies that there exists a similarity
transformation $\mathcal{U}$ such that the Hamiltonian $\mathcal{H}$,
under suitable boundary conditions, can be diagonalized as
\begin{align}
\label{eq:Hbetam}
\mathcal{H}=\sum_{m}h(\beta_{m}) d_{Rm}^{\dagger} d_{Lm},
\end{align}
where $m$ is an index labeling the eigenstates. Here,
$d_{Rm}^\dagger$ and $d_{Lm}^\dagger$ are the right and left creation
operators associated with the right and left eigenstates of
$\mathcal{H}$  \cite{ch.yo.20,he.re.19,gu.yu.21,ya.lu.24}, respectively:
$|R_m\rangle = d_{Rm}^\dagger |0\rangle $,
$|L_m\rangle = d_{Lm}^{\dagger} |0\rangle$.
Due to the non-Hermiticity, $d_{Lm}^\dagger\ne (d_{Rm})^\dagger$, but
they satisfy the biorthogonal anticommutation relation
$\{d_{Lm}, d_{Rn}^\dagger \} = \delta_{mn}$.
The similarity transformation $\mathcal{U}$ is given by
\begin{align}
  \label{eq:cjdm}
  c_j &= \sum_m \mathcal{U}_{jm}   d_{Lm} , \quad
        \mathcal{U}_{jm} = \langle j | R_m \rangle =
        \frac{1}{\sqrt{N}}  e^{i s_{m} j} \nonumber\\
  c_j^\dagger &= \sum_m d_{Rm}^\dagger \mathcal{U}_{mj}^{-1}, \quad
                \mathcal{U}^{-1}_{mj}= \langle L_m | j \rangle = \frac{1}{\sqrt{N}}  e^{-i s_{m} j}
\end{align}
where the generalized momentum $s_m=-i\ln
\beta_m=q_m+i\tau_m$, with $q_m$ and $\tau_m$ being real.
For a general GBZ, the values of $q_m$ are not uniformly distributed
over the interval $[0, 2\pi]$, and $\tau_m$, which corresponds to the
modulus of $\beta_{m}$, is a function of $q_m$ \cite{ya.wa.18,yo.mu.19}.
These make the matrix $\mathcal{U}$ extremely complicated.

For a circular GBZ, the radius $r=|\beta|$ is
constant, and the argument of $\beta$ is uniformly
distributed along the GBZ; see Fig. 1(c) in the main
text. Consequently, $q_m$ coincides with the crystal momentum $k$ in
the conventional Brillouin Zone,
\begin{align}
\label{eq:qmk}
q_m = \frac{2\pi}{N} m, \quad m = 1,2,\cdots,N
\end{align}
with $N$ being the number of unit cells.
In this case, the transformation $\mathcal{U}$ can be simplified as
\begin{align}
  \label{eq:cjdk}
  c_j &= \sum_k \mathcal{U}_{jk}   d_{Lk} , \quad
        \mathcal{U}_{jk} = \frac{1}{\sqrt{N}} r^{j} e^{i k j} \nonumber\\
  c_j^\dagger &= \sum_k d_{Rk}^\dagger \mathcal{U}_{kj}^{-1}, \quad
                \mathcal{U}^{-1}_{kj}= \frac{1}{\sqrt{N}} r^{-j} e^{-i k j},
\end{align}
where the index $m$ is replaced by the index $k$ according to the
relation in~Eq. (\ref{eq:qmk}).

We emphasize that, to obtain the non-Bloch Hamiltonian as in Eq.~(\ref{eq:Hbetam}), one
should impose the proper boundary conditions for the real-space
Hamiltonian $\mathcal{H}$.
For the non-Hermitian SSH model in Eq.~(1) of the main text,
the kinetic term $\mathcal{H}_{K}$ can be written as
\begin{align}
       & \quad\quad \mathcal{H}_{K}= \mathcal{H}_{1}+\mathcal{H}_{2}+\mathcal{H}_{b} \nonumber\\
        \mathcal{H}_{1}= & \sum_{i=1}^{N}\left[\left(t_1+\gamma\right)c_{2i-1}^{\dagger} c_{2i}
                     + \left(t_1-\gamma\right)c_{2i}^{\dagger}
                       c_{2i-1}\right], \nonumber\\
         \mathcal{H}_{2} = & \sum_{i=1}^{N-1}\left[t_2 c_{2i}^{\dagger} c_{2i+1}+t_2
                     c_{2i+1}^{\dagger} c_{2i}\right], \nonumber\\
      \mathcal{H}_b=& \ t_2r^{-N} c_{1}^{\dagger} c_{2N}+t_2r^{N} c_{2N}^{\dagger} c_{1}.
        \label{eq:H_SSHV}
\end{align}
The boundary-condition term $\mathcal{H}_b$ corresponds to neither open nor
periodic boundary conditions. We refer to it as a modified boundary
condition (MBC). The justification for using this boundary condition
is that it is consistent with the non-Bloch Hamiltonian, as shown
below.
%
%
By performing the
similar transformation in Eq.~(\ref{eq:cjdk}) on $\mathcal{H}_{K}$, it is
straightforward to obtain
\begin{align}
\label{eq:HK}
\mathcal{H}_{K} = \sum_{k} \Psi_{Rk}^{\dagger} H(\beta_k) \Psi_{Lk}
\end{align}
in which $\Psi_{Rk}^{\dagger}=(d_{Rk,A}^{\dagger},\ d_{Rk,B}^{\dagger})$,
$\Psi_{Lk}=(d_{Lk,A},\ d_{Lk,B})^T$, and
\begin{equation}
  \label{eq:12}
  H(\beta_{k})=
\begin{pmatrix}
0 & t_1+\gamma + t_2\beta_k^{-1}  \\
t_1-\gamma + t_2 \beta_{k} & 0
\end{pmatrix}
\end{equation}
with $\beta_k=re^{ik}$ and $r=\sqrt{(t_1-\gamma)/(t_1+\gamma)}$. This
non-Bloch Hamiltonian is the starting point of the GBZ theory~\cite{ya.wa.18,yo.mu.19}.

The interacting term $\mathcal{H}_{I}$ of the non-Hermitian SSH model
in Eq.~(1) of the main text can also be transformed into the GBZ basis
using Eq.~(\ref{eq:cjdk}), which yields
\begin{align}
\label{eq:HI}
  \mathcal{H}_{I}&=\sum_{j=1}^{L}V n_{j}n_{j+1}\nonumber\\
                  &= \frac{V}{N} \sum_{k,k',q}(1+e^{iq}) d_{Rk+q,A}^{\dagger}d_{Lk,A}	d_{Rk'-q,B}^{\dagger}d_{Lk',B}	 
\end{align}

\subsection{Construction of a quasi-reciprocal Hamiltonian}

Since the Hamiltonians $\mathcal{H}_{K}$ and $\mathcal{H}_{I}$ in
Eqs. (\ref{eq:HK}) and (\ref{eq:HI}) are indexed by the crystal
momentum $k$, we can perform an Fourier transformation to obtain an
“artificial” quasi-reciprocal Hamiltonian in real space. This idea was
first elaborated in Ref.~\cite{ya.lu.24}. The transformation can be
written as
\begin{align}
  d_{Lk} &= \sum_{j} \mathcal{F}_{kj}^{\dagger}a_j ,\quad  \mathcal{F}_{kj}^{\dagger} = \frac{1}{\sqrt{N}}  e^{-i k j} \nonumber\\
  d_{Rk}^{\dagger} &= \sum_{j} a_j^{\dagger} \mathcal{F}_{jk} ,\quad
                     \mathcal{F}_{jk} = \frac{1}{\sqrt{N}}  e^{i k j}
\label{eq:dkaj}
\end{align}
where $a_{j}^{\dagger}$($a_{j}$) is the creation (annihilation)
operator on the quasi-reciprocal lattice, and the sublattice indices
$A$ and $B$ are not explicitly written.
The final quasi-reciprocal Hamiltonian reads
\begin{align}
\label{eq:Ht_SSHV}
  \tilde{\mathcal{H}}= & \sum_{i=1}^{N}\left[\left(t_1+\gamma\right)a_{2i-1}^{\dagger} a_{2i}
                         +\left(t_1-\gamma\right)a_{2i}^{\dagger} a_{2i-1}\right]\nonumber\\
                       & +\sum_{i=1}^{N}\left[t_2 r a_{2i}^{\dagger}
                         a_{2i+1}+\frac{t_2}{r} a_{2i+1}^{\dagger} a_{2i}\right]
                         +\sum_{j=1}^{L} V\tilde{n}_j\tilde{n}_{j+1},
\end{align}
in which $\tilde{n}_j=a_j^{\dagger}a_j$
is the particle number operator in the new basis.
Compared with the original SSH model in Eq.~(1) of the main
text, the intercell hopping amplitude and the interaction strength
remain unchanged, while the intracell hopping is renormalized.

\begin{figure}[htb]
	\centering
	\includegraphics[width=8.6cm]{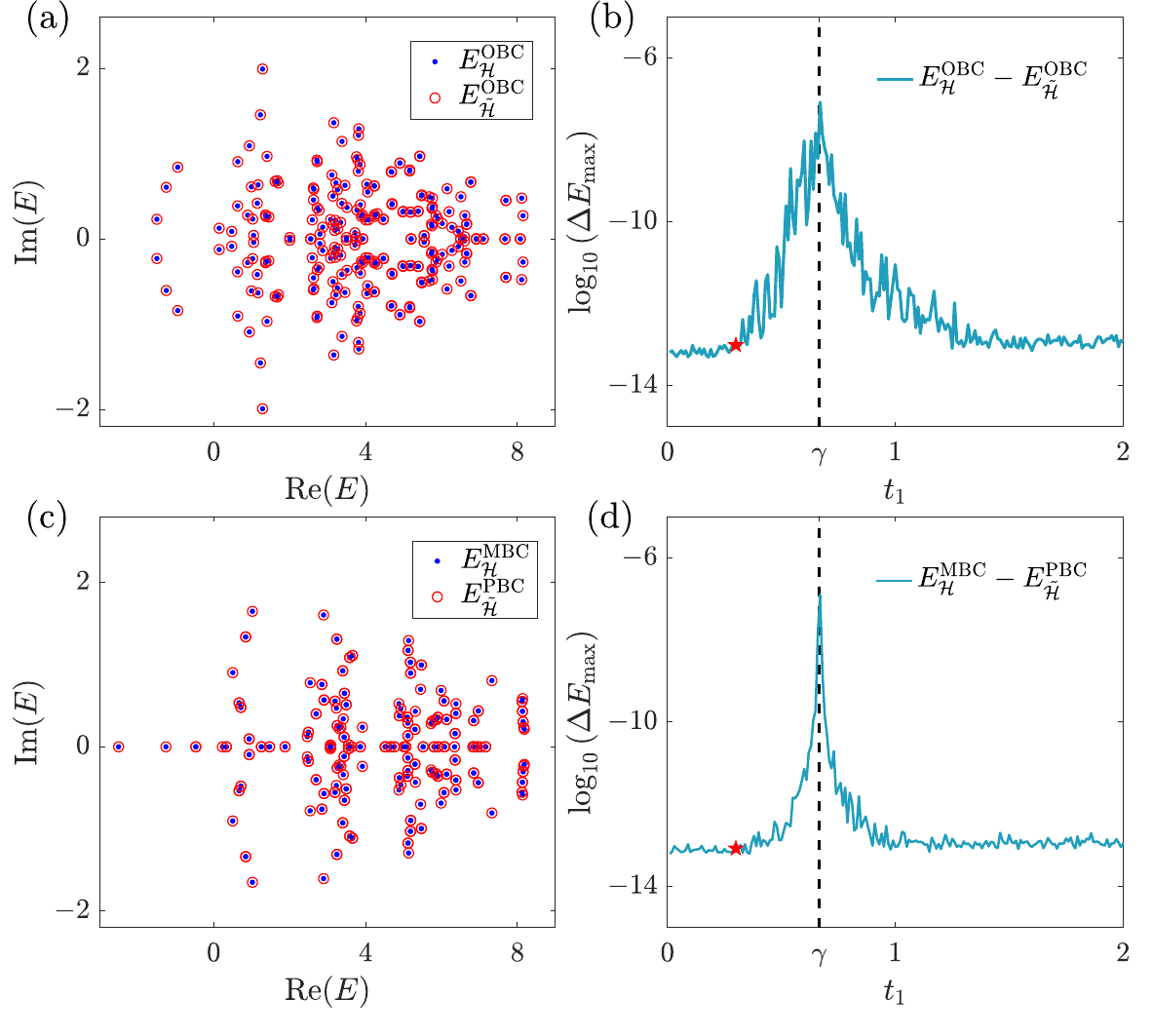}
	\caption{(a,c) Energy spectra of $\mathcal{H}$ (blue dots) and
          $\tilde{\mathcal{H}}$ (red circles) under different boundary
          conditions at $t_1=0.3$.  In panel (a), both
          $\mathcal{H}$ and $\tilde{\mathcal{H}}$ are under OBC. In
          panel (c), $\mathcal{H}$ is under the MBC defined
          in~Eq. (\ref{eq:H_SSHV}), while $\tilde{\mathcal{H}}$ is under
          PBC. (b,d) Maximum absolute difference $\Delta E_{\mathrm{max}}$
          between the energy spectra of $\mathcal{H}$ and
          $\tilde{\mathcal{H}}$ as a function of $t_1$, corresponding to
          $E_{\mathcal{H}}^{OBC}-E_{\tilde{\mathcal{H}}}^{OBC}$ and
           $E_{\mathcal{H}}^{MBC}-E_{\tilde{\mathcal{H}}}^{PBC}$, respectively. The red stars
          mark the data in panels (a) and (c), and the
          dashed line denotes the exceptional point at $t_1 = \gamma$.
          Common parameters: $\gamma=2/3$ and $L=10$.}
	\label{fig:H_Ht_E}
\end{figure}

The quasi-reciprocal Hamiltonian $\mathcal{\tilde{H}}$ is connected to
the original Hamiltonian $\mathcal{H}$ via a similarity transformation
$\mathcal{S}$~\cite{ya.lu.24}. This can be seen by substituting the
transformation in Eq.~(\ref{eq:dkaj}) into Eq.~(\ref{eq:cjdk}), which
yields
\begin{align}
\label{eq:cjaj}
c_{j} &= \sum_{m} \mathcal{S}_{jm}a_m = r^{j}a_{j} , \nonumber\\
c_{j}^{\dagger} &= \sum_{m} a_m^{\dagger}\mathcal{S}_{mj}^{-1} = r^{-j}a_{j}^{\dagger}.
\end{align}
That is, the transformation $\mathcal{S}$ is diagonal:
$\mathcal{S}=\mathcal{U}\mathcal{F}^{\dagger}=\mathrm{diag}\{r,r,r^2,r^2,\dots,r^{N-1},r^{N-1},r^{N},r^{N}\}$.
The similarity transformation
$\bar{\mathcal{S}}=\mathrm{diag}\{1,r,r,r^2,\dots,r^{N-1},r^{N}\}$,
well studied in Ref.~\cite{ya.wa.18}, can be obtained from
$\mathcal{S}$ by an additional similarity transformation
$\mathcal{S}'$, such that $\bar{\mathcal{S}}=\mathcal{S}\mathcal{S}'$,
with $\mathcal{S}'=\mathrm{diag}\{1/r,1,1/r,1,\dots,1/r,1\}$ \cite{ya.lu.24}.
The similarity transformation $\mathcal{S}$ can also be extended to
the many-body basis. Without loss of generality, the $\alpha$-th many-body
basis vector of quasi-reciprocal Hamiltonian $\tilde{\mathcal{H}}$ can
be written as
$|\tilde{\alpha}\rangle=a^{\dagger}_i\cdots a^{\dagger}_j|0 \rangle$.
Using Eq. (\ref{eq:cjaj}), we obtain
\begin{align}
\label{eq:2}
|\tilde{\alpha}\rangle
  =(r^i\cdots r^j)c^{\dagger}_i\cdots c^{\dagger}_j |0\rangle
  =(r^i\cdots r^j) |\alpha \rangle,
\end{align}
where $|\alpha\rangle=c^{\dagger}_i\cdots c^{\dagger}_j|0\rangle$ denotes
the $\alpha$-th many-body basis of the original Hamiltonian $\mathcal{H}$.
Therefore, the similarity transformation matrix $\mathcal{S}_m$ in the
many-body basis is also diagonal:
\begin{align}
\label{eq:1}
(\mathcal{S}_{m})_{\alpha\alpha}=\langle \alpha | \tilde{\alpha} \rangle = r^i\cdots r^j.
\end{align}

In Fig. \ref{fig:H_Ht_E}, we present the numerical energy spectra of
$\mathcal{H}$ and $\tilde{\mathcal{H}}$ under different boundary
conditions.  As shown in panels (a) and (b), the energy spectra of
$\mathcal{H}$ and $\tilde{\mathcal{H}}$ are in exact agreement under
OBC. As shown in panels (c) and (d), the energy spectrum of
$\tilde{\mathcal{H}}$ under PBC coincides with that of $\mathcal{H}$
under the MBC defined in Eq. (\ref{eq:H_SSHV}). The maximum absolute
difference between the energy spectra, shown in
Fig.~\ref{fig:H_Ht_E}(b) and (d), is close to machine precision,
except in the vicinity of the exceptional point at $t_1 = \gamma$.
The behavior of the energy spectra is consistent with our previous
analysis of the similarity transformation.




When the GBZ is non-circular, the similarity transformations defined
in Eq. (\ref{eq:cjdm}) still exist. However, the GBZ radius
$r=|\beta_m|=e^{-\tau_m}$ is no longer a constant but becomes a
function of $q_m$, the argument (angle) of $\beta_m$. Moreover, the
angles $q_m$ are not uniformly distributed over $[0, 2\pi]$. This
makes it difficult to construct a regular quasi-reciprocal Hamiltonian
in real space. If one attempts to perform an inverse Fourier
transformation to the non-Bloch Hamiltonian, both the hopping and
interaction terms inevitably acquire long-range contributions
\cite{ya.lu.24,le.li.20}.  In this case, one cannot write a concise
modified boundary condition such as that in Eq.~(\ref{eq:H_SSHV}), since it would
involve hopping processes of all ranges across the boundary.





\section{density distribution}
\label{App:A}

\begin{figure}[tb]
	\centering
	\includegraphics[width=8.6cm]{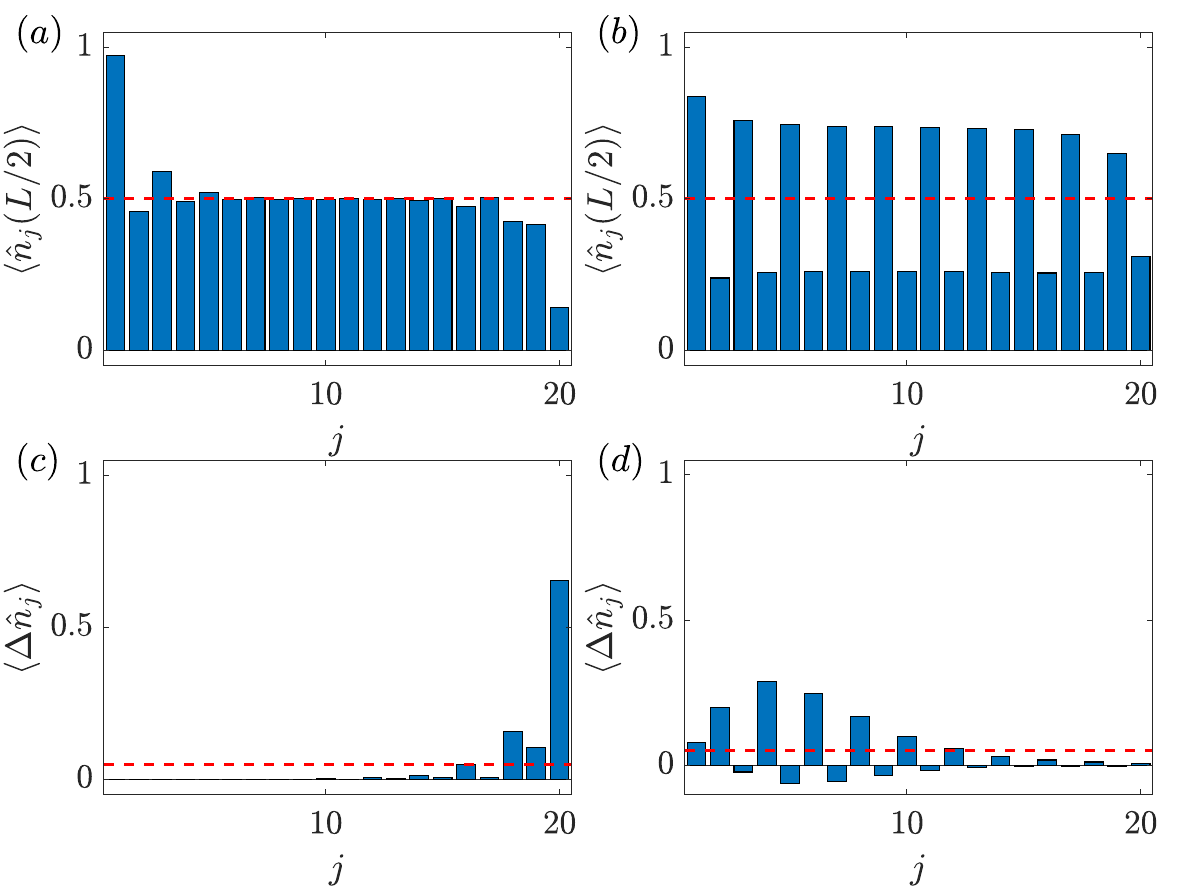}
	\caption{(a,b) particle density distribution at
          half-filling. (c,d) particle density difference
          distribution. Topological phase $(t_1 = 0.3)$ in (a,c);
          trivial phase $(t_1 = 1.5)$ in (b,d). The red dashed lines
          denote the corresponding average values. Common parameters:
          $V=1, L=20$, OBC.  }
	\label{fig:densityV1}
\end{figure}

In this section, we present the density profile of the original
Hamiltonian given in Eq.~(1) of the main text. Using the many-body
right eigenstates, we compute the density at site $j$,
$\rho_{j}=\langle\psi_R|\hat{n}_j|\psi_R\rangle$, as well as the density
difference between the $N$-particle and ($N+1$)-particle systems \cite{zh.pa.23},
\begin{equation}
  \left\langle\Delta \hat{n}_j\right\rangle
  =\langle \psi_R(N+1) | \hat{n}_j |\psi_{R}(N+1)\rangle
  -\langle \psi_R(N) | \hat{n}_j |\psi_{R}(N) \rangle.
	\label{eq:deltan}
\end{equation}
where $N$ represents the total number of particles in the system.
In this paper, we focus on the half-filled case with $N = L/2$.

\begin{figure}[tb]
	\centering
	\includegraphics[width=8.6cm]{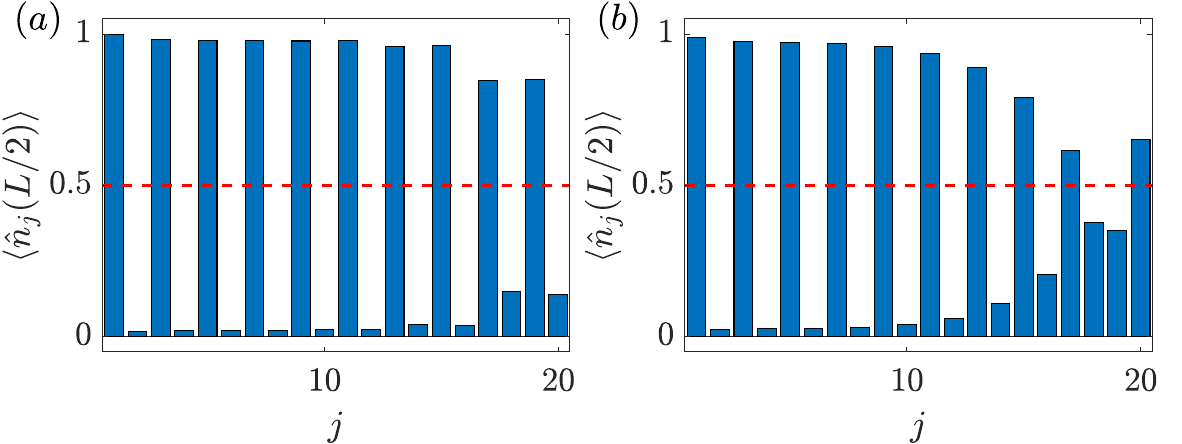}
	\caption{Particle density distribution at half-filling for the
          topological phase with $t_1 = 0.3$ (a) and the topologically
          trivial phase with $t_1 = 1.5$ (b). They all exhibited the
          CDW pattern. The red dashed lines denote the corresponding
          average values. Common parameters: $V=8, L=20$, OBC.}
	\label{fig:densityV8}
\end{figure}

The density distribution for weak interactions $(V=1)$ is shown in
Fig.~\ref{fig:densityV1}. Due to the nonreciprocal hopping, the system
exhibits a many-body skin effect, causing particles to accumulate
preferentially at the left boundary rather than the right one.
In the topological phase [Fig.~\ref{fig:densityV1}(a) and (c)], the
$(N+1)$-th added particle tends to accumulate near the right boundary
due to the presence of edge states. In contrast, in the
topologically trivial phase [Fig.~\ref{fig:densityV1}(b) and (d)], the
$(N+1)$-th added particle is distributed over a broad spatial region.
The density distribution for strong interactions $(V=8)$ is shown in
Fig.~\ref{fig:densityV8}. In both the topological and topologically
trivial phases, a clear CDW pattern is observed, coexisting with the
non-Hermitian skin effect. Notably, a kink in the density profile
appears in Fig.~\ref{fig:densityV8}(b). Such a CDW phase featuring a
kink ($\mathrm{nCDW_k}$) has been investigated in a recent work
Ref.~\cite{zh.pa.25}.


\section{structure factor}
\label{App:B}




For non-Hermitian systems, the structure factor can be defined solely in terms of the right eigenstates as
\begin{align}
\label{eq:SkRR}
S^{RR}(k) &= \frac{1}{L^2} \sum_{i,j} e^{ik(i-j)}
\Big(
     \langle \psi_R | \tilde{n}_i \tilde{n}_j | \psi_R \rangle \nonumber \\
     &\quad - \langle \psi_R | \tilde{n}_i | \psi_R \rangle
              \langle \psi_R | \tilde{n}_j | \psi_R \rangle
\Big),
\end{align}
or, alternatively, using both the left and right eigenstates,
\begin{align}
\label{eq:SkRL}
S^{RL}(k) &= \frac{1}{L^2} \sum_{i,j} e^{ik(i-j)}
\Big(
     \langle \psi_L | \tilde{n}_i \tilde{n}_j | \psi_R \rangle \nonumber \\
     &\quad - \langle \psi_L | \tilde{n}_i | \psi_R \rangle
      \langle \psi_L | \tilde{n}_j | \psi_R \rangle
\Big).
\end{align}

\begin{figure}[htb]
	\centering
	\includegraphics[width=8.6cm]{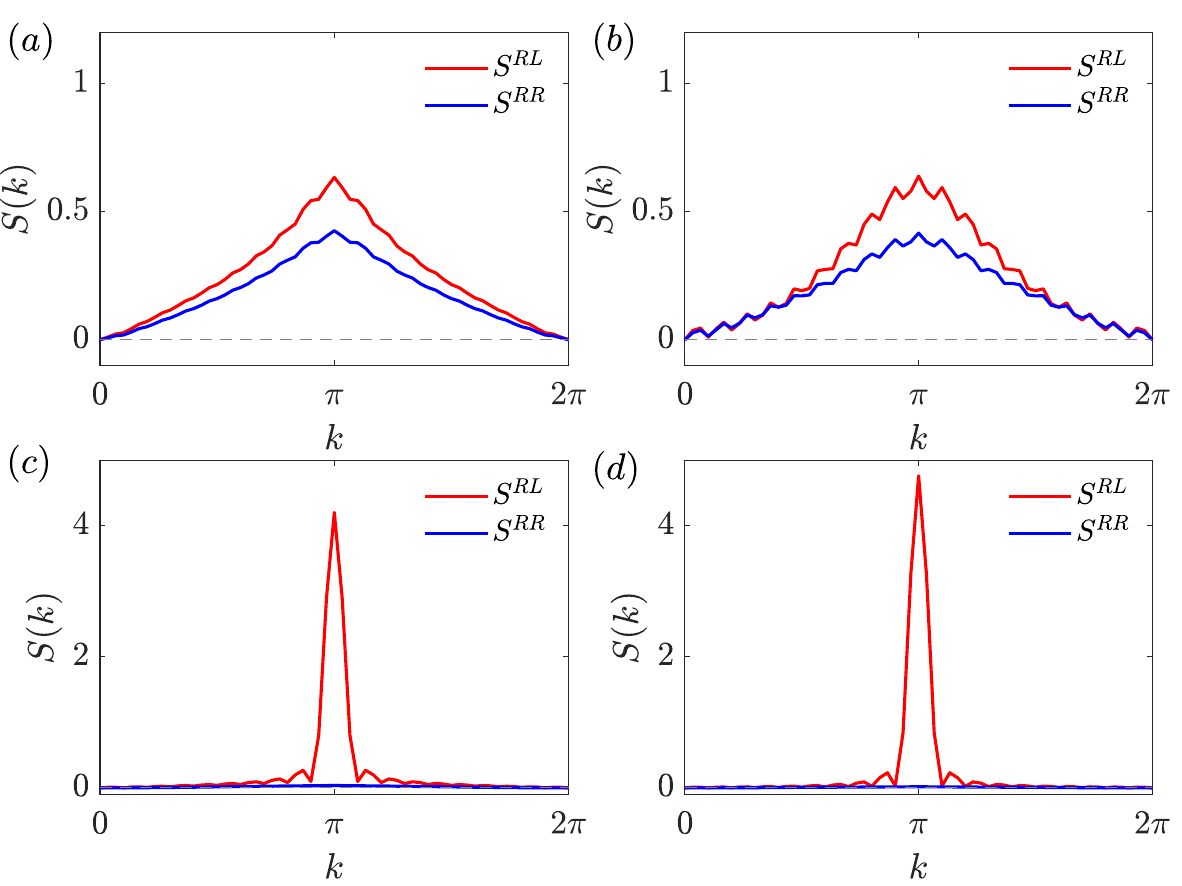}
	\caption{The structural factors $S^{RL}(k)$ and $S^{RR}(k)$
          as functions of $k$ in different phases. (a) Topological
          phase $(t_1=0.3, V=1)$. (b) Trivial phase $(t_1=1.5,
          V=1)$. (c) Topological CDW phase $(t_1=0.3, V=8)$. (d)
          Trivial CDW phase $(t_1=1.5, V=8)$. Common parameters:
          $\gamma=2/3, L=20$, PBC.}
	\label{fig:kScdw}
\end{figure}

For comparison, we present in Fig.~\ref{fig:kScdw} both 
$S^{RL}$ and $S^{RR}$ as functions of 
$k$ in different phases.
As shown in Fig.~\ref{fig:kScdw}(a) and (b), both $S^{RL}$ and
$S^{RR}$ exhibit a broad distribution over $k$ when the interaction is
small ($V=1$), implying that the CDW order is not strong in the system.
When $V=8$ is large, $S^{RL}(k)$ exhibits a sharp peak at $k=\pi$,
indicating a CDW state, whereas $S^{RR}(k)$ remains nearly zero; see
Fig.~\ref{fig:kScdw}(c) and (d).
The biorthogonal structure factor $S^{RL}(k)$ is consistent with the
density profile presented in Fig.~\ref{fig:densityV8}, and therefore
is the proper characterization for the non-Hermitian CDW phase.


\section{CDW phase boundary}
\label{App:CDWphase}

\subsection{Fitted phase boundary of the CDW phase in the Hermitian SSH model}

First, we illustrate the fitting procedure for the
Hermitian SSH model with nearest-neighbor interactions. The
Hamiltonian reads
\begin{align}
\label{eq:H_SSHVt}
\mathcal{H}(t_1,t_2,V) =& \sum_{m=1}^{N}\left( t_1 c_{m,A}^{\dagger} c_{m,B}
                          +t_2 c_{m,B}^{\dagger} c_{m+1,A}+H . c .\right)\nonumber\\
		&+\sum_{j=1}^{L} Vn_jn_{j+1},
\end{align}
Under PBC, the Hamiltonian (\ref{eq:H_SSHVt})
remains invariant when exchanging the hopping parameters $t_1$ and
$t_2$, \textit{i.e.},
\begin{equation}
  \label{eq:HH}
  \mathcal{H}(t_1,t_2,V) =\mathcal{H}(t_2,t_1,V),
\end{equation}
since this exchange
is equivalent to a relabeling of the lattice sites \cite{si.ma.14}.
Furthermore, for the ground state energy $E_0$ and wave function
$\left| \psi_{0}\right\rangle$, one has
\begin{equation}
\label{eq:H/c}
\mathcal{H}\left( \frac{t_1}{c},\frac{t_2}{c},\frac{V}{c} \right)
\left| \psi_0\right\rangle = \frac{E_0}{c} \left| \psi_0\right\rangle,
\end{equation}
where $c$ is a positive real constant.
In this paper, we set $t_2=1$ as the energy unit. From
Eqs. (\ref{eq:HH}) and (\ref{eq:H/c}), we can derive
\begin{align}
\label{eq:Ht1H}
\mathcal{H}(t_1,1,V)=t_1\mathcal{H}\left( 1,\frac{1}{t_1},\frac{V}{t_1}\right)
		=t_1\mathcal{H}\left( \frac{1}{t_1},1,\frac{V}{t_1}\right).
\end{align}
The above equation shows that the properties of $\mathcal{H}$ with
parameters $(t_1,V)$ correspond to those with parameters
$(1/t_1,V/t_1)$.
Assume that the CDW phase boundary $V_c$ in the regime $t_1 > 1$ takes
the following functional form:
\begin{equation}
  V_c=f(t_1), \qquad  (0<t_1<1)
\end{equation}
Note that multiplying $\mathcal{H}$ by a constant factor $t_1$ does
not affect the CDW structure factor $S(\pi)$, nor the CDW phase
boundary.  According to Eq.~(\ref{eq:Ht1H}), we can perform a
transformation $(t_1,V) \rightarrow (1/t_1,V/t_1)$, which yields
\begin{equation}
   \frac{V_c}{t_1}=f(\frac{1}{t_1}), \qquad  (t_1>1)
 \end{equation}
Consequently, using the CDW phase boundary in the regime $0<t_1<1$, one
can map out the phase boundary in the regime
$t_{1}>1$, which is given by
\begin{equation}
	V_c=t_1f(\frac{1}{t_1}), \qquad  (t_1>1).
	\label{eq:Vc}
\end{equation}

\begin{figure}[htb]
	\centering
	\includegraphics[width=1.0\linewidth]{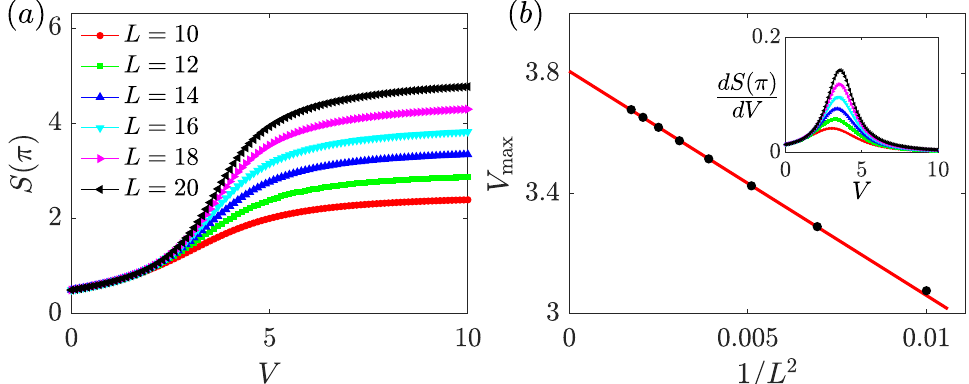}
	\caption{(a) CDW structure factor $S(\pi)$ as a function of
          interaction $V$ for various lattice sizes $L$. (b)
          Extrapolation of $V_{\mathrm{max}}$ as a function of
          $1/L^2$. The inset shows the derivative of structure factor
          $d S(\pi)/dV$ for the same parameters as in (a).
          $V_{\mathrm{max}}$ is the point at which the derivative
          $d S(\pi)/dV$ is maximized. Calculations are performed for
          the Hermitian SSH model under PBC with $t_1=0.5$.}
	\label{fig:ScdwH}
\end{figure}




The critical interaction $V_c$ for the CDW phase transition is
determined numerically using the same procedure described in the main text:
locating the maximum of $\frac{d S(\pi)}{d V}$ and then extrapolating
the corresponding $V_{\mathrm{max}}$ to the thermodynamic limit.
Figure~\ref{fig:ScdwH} shows the structure factor $S(\pi)$ and the
corresponding maximum
position $V_{\mathrm{max}}$ for Hermitian SSH model with $t_1 =
0.5$. By extrapolating $V_{\mathrm{max}}$, the critical interaction is
determined to be $V_c = 3.8$.


\begin{figure}[htb]
	\centering
	\includegraphics[width=1.0\linewidth]{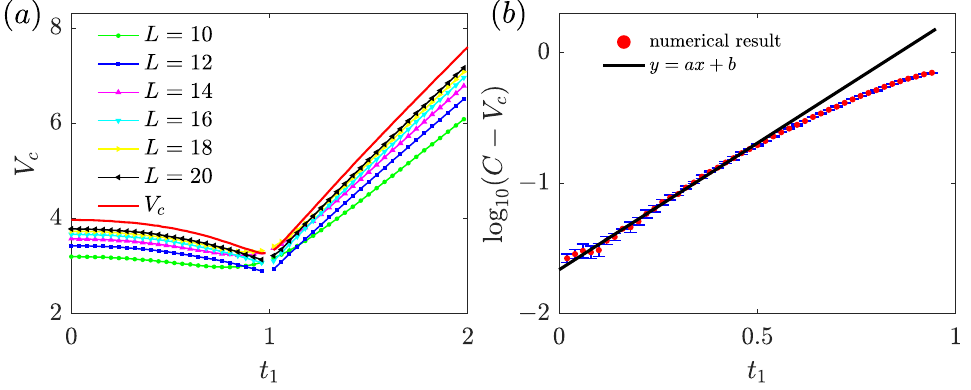}
	\caption{(a) Extrapolated critical interaction $V_c$ as a
          function of $t_1$. (b) $\log_{10}(C-V_c)$ plotted versus
          $t_1$, with $a=2$, $b=-1.699$, and $C=4$.  The red points
          denote the ED results, and the error bars indicate the
          standard deviation obtained from five datasets.
          Calculations are performed for the Hermitian SSH model under
          PBC.}
	\label{fig:plotVc}
\end{figure}

We present the numerically extrapolated critical interaction $V_c$ for
different values of $t_1$ in Fig.~\ref{fig:plotVc}(a); See the red
curve. In the topological regime $0<t_1<1$, $V_c$ decreases with
increasing $t_1$, whereas in the trivial regime $t_1>1$, $V_c$
increases roughly linearly with $t_1$ \cite{me.ju.23}.  The
correlation length $\xi$ of the Hermitian SSH model diverges at the
topological phase transition point $t_1=1$ \cite{si.ma.14}.  Numerical
results obtained using ED method are reliable when the correlation
length is much smaller than the system size $(\xi\ll L)$,
\textit{i.e.}, sufficiently far from the critical point
($t_1\rightarrow 0$ or $t_1\rightarrow \infty$). In the vicinity of
$t_1=1$, numerical methods become unreliable due to strong finite-size
effects.  Fortunately, at the isotropic point $t_1=t_2=1$, the
Hamiltonian in Eq.~(\ref{eq:H_SSHVt}) is exactly solvable using the
Bethe ansatz, which yields a critical interaction
$V_{c}=2$~\cite{ca.ci.11,mi.ca.11}.

Assuming that $V_{c}$ decreases exponentially in the regime
$0 < t_1 < 1$, we fit the phase boundary using the formula
$V_c = C - 10^{at_1+b}$.  By imposing the exact values of $V_c$ at the
limits of $t_{1}=1$ and $t_1=0$, the parameters are determined to be
$a=2$, $b=-1.699$, and $C=4$.
In Fig.~\ref{fig:plotVc}(b), we plot $\log_{10}(C-V_c)$ as a function
of $t_1$. The ED results exactly match the fitting black line in the regime
$0<t_1<0.5$, but deviate from it in the regime $0.5<t_1<1$ due to
strong finite-size effects.
The phase boundary in the regime $t_1 > 1$ can be mapped out through
Eq.~(\ref{eq:Vc}); therefore we obtain
\begin{align}\label{eq:Vc_H}
	V_c =
	\begin{cases} 
			4-\frac{10^{2t_1}}{50} & \quad 0<t_1<1, \\
			4t_1-\frac{10^{2/t_1}}{50}t_1 & \quad t_1>1.
	\end{cases}
\end{align}

The phase boundary can also be expressed in terms of dimerization
strength $\delta=(t_2-t_1)/2$. Given $t_1=1-\delta$ and
$t_2=1+\delta$, the critical interaction takes the form
\begin{equation}
	V_c(\delta) = 4(1+|\delta|)-\frac{(1+|\delta|)}{50} 10^{\frac{2-2|\delta|}{1+|\delta|}},\ \ |\delta|<1.
	\label{eq:Vc_deltat}
\end{equation}
The phase boundary given in Eq.~(\ref{eq:Vc_deltat}) is
in good agreement with the recent numerical results reported in
Ref. \cite{ve.la.25}.

\subsection{Fitted CDW phase boundary of the non-Hermitian SSH model
  with $t_{1}>\gamma$}

Next, we study the CDW phase boundary of the quasi-reciprocal SSH
model defined in Eq. (\ref{eq:Ht_SSHV}). When $t_{1}>\gamma$, this
quasi-reciprocal model $\tilde{\mathcal{H}}$ can be mapped onto a Hermitian SSH model
$\bar{\mathcal{H}}$ via the similarity transformation $\mathcal{S}'=\mathrm{diag}\{1/r,1,1/r,1,\dots,1/r,1\}$
\cite{ya.lu.24}. The fermionic operators transform as
\begin{align}
\label{eq:aiei}
  a_{i,A} & = \frac{1}{r}e_{i,A}, \quad a_{i,B} = e_{i,B}\nonumber\\
  a_{i,A}^{\dagger} & = r e_{i,A}^{\dagger}, \quad a_{i,B}^{\dagger} = e_{i,B}^{\dagger}
\end{align}
After the transformation, the Hamiltonian 
$\bar{\mathcal{H}}$ takes the form
\begin{align}
\label{eq:Hb_SSHV}
  \bar{\mathcal{H}}= \sum_{i=1}^{N}\left[\bar{t}_{1}e_{2i-1}^{\dagger} e_{2i}
                         +\bar{t}_2e_{2i}^{\dagger} e_{2i+1} +
                       \mathrm{H.c.} \right]
                       +\sum_{j=1}^{L} V\bar{n}_j\bar{n}_{j+1},
\end{align}
where the intracell hopping amplitude is
$\bar{t}_1 = \sqrt{(t_1 - \gamma)(t_1 + \gamma)}$, the intercell
hopping amplitude is $\bar{t}_{2}=t_{2}$, and
$\bar{n}_{j}=e_{j}^{\dagger} e_{j}$ denotes the particle number
operator~\cite{ya.wa.18}.

Under PBC, the real-space matrices of $\tilde{\mathcal{H}}$ and
$\bar{\mathcal{H}}$ are related by
\begin{equation}
  \bar{H}_{\mathrm{PBC}} =(\mathcal{S}'_m)^{-1}\tilde{H}_{\mathrm{PBC}}\mathcal{S}'_{m},
\end{equation}
where $\mathcal{S}'_m$ is the diagonal similarity transformation
in the many-body basis associated with $\mathcal{S}'$.
We denote the left and right eigenstates of
$\tilde{H}_{\mathrm{PBC}}$ as $|\psi_L\rangle$ and $|\psi_R\rangle$,
respectively, and the eigenstate of $\bar{H}_{\mathrm{PBC}}$ as
$|\bar{\psi}\rangle$, which can be expanded on many-body basis as
$|\bar{\psi}\rangle=\sum_{\alpha}C_{\alpha}|\bar{\alpha}\rangle$.
These states satisfy $|\psi_R\rangle=\mathcal{S}'_m|\bar{\psi}\rangle$
and $\langle\psi_L|=\langle\bar{\psi}|(\mathcal{S}'_m)^{-1}$. Since
$\mathcal{S}'_m$ is diagonal, it follows that
\begin{align}
\label{eq:3}
\langle \psi_L | \tilde{n}_j | \psi_R \rangle =&
   \langle\bar\psi |(\mathcal{S}'_{m})^{-1} \bar{n}_j
   \mathcal{S}'_{m}|\bar\psi\rangle\nonumber\\
  =&\sum_{\alpha\beta}
     C_\alpha^{*}C_\beta(\mathcal{S}'_{m})_{\alpha\alpha}^{-1}
     (\mathcal{S}'_{m})_{\beta\beta} \langle \bar\alpha
     |\bar{n}_j|\bar\beta\rangle \nonumber\\
  =& \sum_{\alpha}|C_\alpha|^{2}\langle \bar\alpha|
     \bar{n}_j|\bar\alpha\rangle
     =\langle \bar\psi | \bar{n}_j | \bar\psi \rangle.
\end{align}
That is, in the biorthogonal basis, the density distributions for
$\tilde{H}_{\mathrm{PBC}}$ and $\bar{H}_{\mathrm{PBC}}$ are
identical. Similarly, one has
$\langle \psi_L | \tilde{n}_i\tilde{n}_j | \psi_R \rangle = \langle
\bar\psi | \bar{n}_{i}\bar{n}_j | \bar{\psi} \rangle$.
Therefore, according to Eq.~(\ref{eq:SkRL}), the structure factor
$S^{RL}(\pi)$ of $\tilde{H}_{\mathrm{PBC}}$ is identical to
that of $\bar{H}_{\mathrm{PBC}}$. In the regime
$t_1>\gamma$, we determine the CDW phase boundary of the
quasi-reciprocal model $\tilde{\mathcal{H}}$ from the structure
factor of its Hermitian counterpart $\bar{\mathcal{H}}$.
Explicitly, we first determine the CDW phase boundary $V_{c}(\bar{t}_{1})$ of
the Hermitian $\bar{\mathcal{H}}$, and then replace $\bar{t}_{1}$ with
$\sqrt{(t_1 - \gamma)(t_1 + \gamma)}$ to obtain the CDW phase boundary
of $\tilde{\mathcal{H}}$.




%